\documentclass[12pt,a4paper,reqno]{article}
\usepackage{amsmath,amsthm,amsfonts}

\def\LL{\leavevmode\setbox0=\hbox{L}\hbox to\wd0{\hss\char'40L}}

\def\la{\lambda}
\def\rh{\rho}

\def\today{\ifcase\month\or
 January\or February\or March\or April\or May\or June\or
 July\or August\or September\or October\or November\or December\fi
 \space\number\day, \number\year}

\def\({\left(}
\def\){\right)}
\def\[{\left[}
\def\]{\right]}

\def\3{\ss}

\catcode`\@=11

\thinlines
\newskip\Einheit \Einheit=0.6cm
\newcount\xcoord \newcount\ycoord
\newdimen\xdim \newdimen\ydim \newdimen\PfadD@cke \newdimen\Pfadd@cke
\def\PfadDicke#1{\PfadD@cke#1 \divide\PfadD@cke by2 \Pfadd@cke\PfadD@cke \multiply\PfadD@cke by2}
\long\def\LOOP#1\REPEAT{\def\BODY{#1}\ITERATE}
\def\ITERATE{\BODY \let\next\ITERATE \else\let\next\relax\fi \next}
\let\REPEAT=\fi
\def\Punkt{\hbox{\raise-2pt\hbox to0pt{\hss\scriptsize$\bullet$\hss}}}
\def\DuennPunkt(#1,#2){\unskip
  \raise#2 \Einheit\hbox to0pt{\hskip#1 \Einheit
          \raise-2.5pt\hbox to0pt{\hss\normalsize$\bullet$\hss}\hss}}
\def\NormalPunkt(#1,#2){\unskip
  \raise#2 \Einheit\hbox to0pt{\hskip#1 \Einheit
          \raise-3pt\hbox to0pt{\hss\large$\bullet$\hss}\hss}}
\def\DickPunkt(#1,#2){\unskip
  \raise#2 \Einheit\hbox to0pt{\hskip#1 \Einheit
          \raise-4pt\hbox to0pt{\hss\Large$\bullet$\hss}\hss}}
\def\Kreis(#1,#2){\unskip
  \raise#2 \Einheit\hbox to0pt{\hskip#1 \Einheit
          \raise-4pt\hbox to0pt{\hss\Large$\circ$\hss}\hss}}
\def\Diagonale(#1,#2)#3{\unskip\leavevmode
  \xcoord#1\relax \ycoord#2\relax
      \raise\ycoord \Einheit\hbox to0pt{\hskip\xcoord \Einheit
         \unitlength\Einheit
         \line(1,1){#3}\hss}}
\def\AntiDiagonale(#1,#2)#3{\unskip\leavevmode
  \xcoord#1\relax \ycoord#2\relax \advance\xcoord by -0.05\relax
      \raise\ycoord \Einheit\hbox to0pt{\hskip\xcoord \Einheit
         \unitlength\Einheit
         \line(1,-1){#3}\hss}}
\def\Pfad(#1,#2),#3\endPfad{\unskip\leavevmode
  \xcoord#1 \ycoord#2 \thicklines\ZeichnePfad#3\endPfad\thinlines}
\def\ZeichnePfad#1{\ifx#1\endPfad\let\next\relax
  \else\let\next\ZeichnePfad
    \ifnum#1=1
      \raise\ycoord \Einheit\hbox to0pt{\hskip\xcoord \Einheit
         \vrule height\Pfadd@cke width1 \Einheit depth\Pfadd@cke\hss}%
      \advance\xcoord by 1
    \else\ifnum#1=2
      \raise\ycoord \Einheit\hbox to0pt{\hskip\xcoord \Einheit
        \hbox{\hskip-1pt\vrule height1 \Einheit width\PfadD@cke depth0pt}\hss}%
      \advance\ycoord by 1
    \else\ifnum#1=3
      \raise\ycoord \Einheit\hbox to0pt{\hskip\xcoord \Einheit
         \unitlength\Einheit
         \line(1,1){1}\hss}
      \advance\xcoord by 1
      \advance\ycoord by 1
    \else\ifnum#1=4
      \raise\ycoord \Einheit\hbox to0pt{\hskip\xcoord \Einheit
         \unitlength\Einheit
         \line(1,-1){1}\hss}
      \advance\xcoord by 1
      \advance\ycoord by -1
    \fi\fi\fi\fi
  \fi\next}
\def\hSSchritt{\leavevmode\raise-.4pt\hbox to0pt{\hss.\hss}\hskip.2\Einheit
  \raise-.4pt\hbox to0pt{\hss.\hss}\hskip.2\Einheit
  \raise-.4pt\hbox to0pt{\hss.\hss}\hskip.2\Einheit
  \raise-.4pt\hbox to0pt{\hss.\hss}\hskip.2\Einheit
  \raise-.4pt\hbox to0pt{\hss.\hss}\hskip.2\Einheit}
\def\vSSchritt{\vbox{\baselineskip.2\Einheit\lineskiplimit0pt
\hbox{.}\hbox{.}\hbox{.}\hbox{.}\hbox{.}}}
\def\DSSchritt{\leavevmode\raise-.4pt\hbox to0pt{%
  \hbox to0pt{\hss.\hss}\hskip.2\Einheit
  \raise.2\Einheit\hbox to0pt{\hss.\hss}\hskip.2\Einheit
  \raise.4\Einheit\hbox to0pt{\hss.\hss}\hskip.2\Einheit
  \raise.6\Einheit\hbox to0pt{\hss.\hss}\hskip.2\Einheit
  \raise.8\Einheit\hbox to0pt{\hss.\hss}\hss}}
\def\dSSchritt{\leavevmode\raise-.4pt\hbox to0pt{%
  \hbox to0pt{\hss.\hss}\hskip.2\Einheit
  \raise-.2\Einheit\hbox to0pt{\hss.\hss}\hskip.2\Einheit
  \raise-.4\Einheit\hbox to0pt{\hss.\hss}\hskip.2\Einheit
  \raise-.6\Einheit\hbox to0pt{\hss.\hss}\hskip.2\Einheit
  \raise-.8\Einheit\hbox to0pt{\hss.\hss}\hss}}
\def\SPfad(#1,#2),#3\endSPfad{\unskip\leavevmode
  \xcoord#1 \ycoord#2 \ZeichneSPfad#3\endSPfad}
\def\ZeichneSPfad#1{\ifx#1\endSPfad\let\next\relax
  \else\let\next\ZeichneSPfad
    \ifnum#1=1
      \raise\ycoord \Einheit\hbox to0pt{\hskip\xcoord \Einheit
         \hSSchritt\hss}%
      \advance\xcoord by 1
    \else\ifnum#1=2
      \raise\ycoord \Einheit\hbox to0pt{\hskip\xcoord \Einheit
        \hbox{\hskip-2pt \vSSchritt}\hss}%
      \advance\ycoord by 1
    \else\ifnum#1=3
      \raise\ycoord \Einheit\hbox to0pt{\hskip\xcoord \Einheit
         \DSSchritt\hss}
      \advance\xcoord by 1
      \advance\ycoord by 1
    \else\ifnum#1=4
      \raise\ycoord \Einheit\hbox to0pt{\hskip\xcoord \Einheit
         \dSSchritt\hss}
      \advance\xcoord by 1
      \advance\ycoord by -1
    \fi\fi\fi\fi
  \fi\next}
\def\Koordinatenachsen(#1,#2){\unskip
 \hbox to0pt{\hskip-.5pt\vrule height#2 \Einheit width.5pt depth1 \Einheit}%
 \hbox to0pt{\hskip-1 \Einheit \xcoord#1 \advance\xcoord by1
    \vrule height0.25pt width\xcoord \Einheit depth0.25pt\hss}}
\def\Koordinatenachsen(#1,#2)(#3,#4){\unskip
 \hbox to0pt{\hskip-.5pt \ycoord-#4 \advance\ycoord by1
    \vrule height#2 \Einheit width.5pt depth\ycoord \Einheit}%
 \hbox to0pt{\hskip-1 \Einheit \hskip#3\Einheit
    \xcoord#1 \advance\xcoord by1 \advance\xcoord by-#3
    \vrule height0.25pt width\xcoord \Einheit depth0.25pt\hss}}
\def\Gitter(#1,#2){\unskip \xcoord0 \ycoord0 \leavevmode
  \LOOP\ifnum\ycoord<#2
    \loop\ifnum\xcoord<#1
      \raise\ycoord \Einheit\hbox to0pt{\hskip\xcoord \Einheit\Punkt\hss}%
      \advance\xcoord by1
    \repeat
    \xcoord0
    \advance\ycoord by1
  \REPEAT}
\def\Gitter(#1,#2)(#3,#4){\unskip \xcoord#3 \ycoord#4 \leavevmode
  \LOOP\ifnum\ycoord<#2
    \loop\ifnum\xcoord<#1
      \raise\ycoord \Einheit\hbox to0pt{\hskip\xcoord \Einheit\Punkt\hss}%
      \advance\xcoord by1
    \repeat
    \xcoord#3
    \advance\ycoord by1
  \REPEAT}
\def\Label#1#2(#3,#4){\unskip \xdim#3 \Einheit \ydim#4 \Einheit
  \def\lo{\advance\xdim by-.5 \Einheit \advance\ydim by.5 \Einheit}%
  \def\llo{\advance\xdim by-.25cm \advance\ydim by.5 \Einheit}%
  \def\loo{\advance\xdim by-.5 \Einheit \advance\ydim by.25cm}%
  \def\o{\advance\ydim by.25cm}%
  \def\ro{\advance\xdim by.5 \Einheit \advance\ydim by.5 \Einheit}%
  \def\rro{\advance\xdim by.25cm \advance\ydim by.5 \Einheit}%
  \def\roo{\advance\xdim by.5 \Einheit \advance\ydim by.25cm}%
  \def\l{\advance\xdim by-.30cm}%
  \def\r{\advance\xdim by.30cm}%
  \def\lu{\advance\xdim by-.5 \Einheit \advance\ydim by-.6 \Einheit}%
  \def\llu{\advance\xdim by-.25cm \advance\ydim by-.6 \Einheit}%
  \def\luu{\advance\xdim by-.5 \Einheit \advance\ydim by-.30cm}%
  \def\u{\advance\ydim by-.30cm}%
  \def\ru{\advance\xdim by.5 \Einheit \advance\ydim by-.6 \Einheit}%
  \def\rru{\advance\xdim by.25cm \advance\ydim by-.6 \Einheit}%
  \def\ruu{\advance\xdim by.5 \Einheit \advance\ydim by-.30cm}%
  #1\raise\ydim\hbox to0pt{\hskip\xdim
     \vbox to0pt{\vss\hbox to0pt{\hss$#2$\hss}\vss}\hss}%
}
\catcode`\@=12

\font\scalefont = cmti8

\thicklines

\newcount\xmin                
\newcount\xmax
\newcount\ymin
\newcount\ymax

\newcount\gridwidth
\newcount\gridheight

\newcount\nofxpoints          
\newcount\nofypoints          

\newcount\dcnta               
\newcount\dcntb               

\def\begingrid#1#2#3#4{                 
        \global\xmin = #1
        \global\ymin = #2
        \global\xmax = #3
        \global\ymax = #4
        \ifnum\xmin > \xmax\errmessage{PATHS: \xmin > \xmax|}\fi
        \ifnum\ymin > \ymax\errmessage{PATHS: \ymin > \ymax|}\fi
        \gridwidth = \xmax
        \gridheight = \ymax
        \advance\gridwidth by -\xmin
        \advance\gridheight by -\ymin
        \nofxpoints = \gridwidth
        \advance\nofxpoints by 1
        \nofypoints = \gridheight
        \advance\nofypoints by 1
        \advance\gridwidth by 4
        \advance\gridheight by 4
        \advance\xmin by -2
        \advance\ymin by -2
        \begin{picture}(\gridwidth, \gridheight)(\xmin, \ymin)
        \advance\gridwidth by -4
        \advance\gridheight by -4
        \advance\xmin by 2
        \advance\ymin by 2
        \global\dcnta = \ymin
        \loop\ifnum\dcnta<\ymax
                \makegridline\dcnta     
                \global\advance\dcnta by 1
        \repeat
        \makegridline\ymax
}

\def\makegridline#1{
   \begingroup
   \global\dcntb = \xmin
   \loop\ifnum\dcntb<\xmax
      \put(\dcntb, #1){\circle*{0.1}}
      \global\advance\dcntb by 1
   \repeat
   \put(\xmax, #1){\circle*{0.1}}
   \endgroup
}

\def\makexaxis{\thinlines
        \dcnta = \gridwidth
        \advance\dcnta by 1
        \put(\xmin, 0){\vector(1,0){\dcnta}}
        \thicklines
}
\def\makeyaxis{\thinlines
        \dcntb = \gridheight
        \advance\dcntb by 1
        \put(0, \ymin){\vector(0,1){\dcntb}}
        \thicklines
}
\def\makeaxes{\thinlines
        \dcnta = \gridwidth
        \advance\dcnta by 1
        \put(\xmin, 0){\vector(1,0){\dcnta}}
        \dcntb = \gridheight
        \advance\dcntb by 1
        \put(0, \ymin){\vector(0,1){\dcntb}}
        \thicklines
}

\def\makexscale{
   \dcnta = \xmin
        \ifnum\dcnta<-9\dcnta=-9\fi
   \loop\ifnum\dcnta<0
                \put(\dcnta,0){\line(0,-1){0.2}}
      \put(\dcnta,-0.5){\makebox(0,0){
                {\scalefont\number\dcnta}}}
      \advance\dcnta by 1
  \repeat
  \dcnta = 1
  \dcntb=\xmax
  \ifnum\dcntb>9\dcntb=9\fi
  \loop\ifnum\dcnta<\dcntb
      \put(\dcnta,0){\line(0,-1){0.2}}
      \put(\dcnta, -0.5){\makebox(0,0){
                {\scalefont\number\dcnta}}}
      \advance\dcnta by 1
  \repeat
  \put(\dcntb,0){\line(0,-1){0.2}}
  \put(\dcntb, -0.4){\makebox(0,0){{\scalefont\number\dcntb}}}
}
\def\makeyscale{
   \dcnta = \ymin
   \loop\ifnum\dcnta<0
      \put(0,\dcnta){\line(-1,0){0.2}}
      \put(-0.4, \dcnta){\makebox(0,0){
                {\scalefont\number\dcnta}}}
      \advance\dcnta by 1
   \repeat
   \dcnta = 1
   \loop\ifnum\dcnta<\ymax
      \put(0,\dcnta){\line(-1,0){0.2}}
      \put(-0.4, \dcnta){\makebox(0,0){
                {\scalefont\number\dcnta}}}
      \advance\dcnta by 1
   \repeat
   \put(-0.4, \ymax){\makebox(0,0){{\scalefont\number\ymax}}}
}

\newsavebox{\skewdowndottedbox}
\newsavebox{\skewupdottedbox}
\savebox{\skewdowndottedbox}(1,1)[lb]{
        \put(0.2,-0.2){\circle*{0.01}}
   \put(0.4,-0.4){\circle*{0.01}}
        \put(0.6,-0.6){\circle*{0.01}}
        \put(0.8,-0.8){\circle*{0.01}}
}
\savebox{\skewupdottedbox}(1,1)[lb]{
        \put(0.2,0.2){\circle*{0.01}}
   \put(0.4,0.4){\circle*{0.01}}
        \put(0.6,0.6){\circle*{0.01}}
        \put(0.8,0.8){\circle*{0.01}}
}

\newsavebox{\hplainbox}
\savebox{\hplainbox}{\put(0,0){\line(1,0){1}}}
\newsavebox{\hdottedbox}
\savebox{\hdottedbox}(1,0)[bl]{
        \put(0.2,0){\circle*{0.01}}
        \put(0.4,0){\circle*{0.01}}
        \put(0.6,0){\circle*{0.01}}
        \put(0.8,0){\circle*{0.01}}
}
\newsavebox{\vplainbox}
\savebox{\vplainbox}{\put(0,0){\line(0,1){1}}}
\newsavebox{\vdottedbox}
\savebox{\vdottedbox}(0,1)[bl]{
        \put(0,0.2){\circle*{0.01}}
        \put(0,0.4){\circle*{0.01}}
        \put(0,0.6){\circle*{0.01}}
        \put(0,0.8){\circle*{0.01}}
}
\newsavebox{\vplaindownbox}
\savebox{\vplaindownbox}{\put(0,0){\line(0,-1){1}}}
\newsavebox{\vdotteddownbox}
\savebox{\vdotteddownbox}(0,1)[bl]{
        \put(0,-0.2){\circle*{0.01}}
        \put(0,-0.4){\circle*{0.01}}
        \put(0,-0.6){\circle*{0.01}}
        \put(0,-0.8){\circle*{0.01}}
}
\newsavebox{\skewupbox}
\savebox{\skewupbox}{\put(0,0){\line(1,1){1}}}
\newsavebox{\skewdownbox}
\savebox{\skewdownbox}{\put(0,0){\line(1,-1){1}}}
\newsavebox{\hstep}             
\newsavebox{\vstep}             
\newsavebox{\sstep}             
\newcount\updownincrement

\def\dosteplist{\afterassignment\handlenextstep\let\next=}
\def\handlenextstep{
        \ifx\next\endList
        \let\next=\relax
        \else
                \ifx\next-
                        \put(\dcnta,\dcntb){\usebox{\hstep}}
                        \advance\dcnta by 1
                \else
                        \ifx\next|
                                \put(\dcnta,\dcntb){\usebox{\vstep}}
                                \advance\dcntb by\updownincrement
                        \else
                                \ifx\next/
                                        \put(\dcnta,\dcntb){\usebox{\sstep}}
                                        \advance\dcnta by 1
                                        \advance\dcntb by\updownincrement
                                \else
                                        \errmessage{PATHS: Wrong symbol path argument.}
                                \fi
                        \fi
                \fi
                \let\next=\dosteplist
        \fi
        \next
}

\def\endgrid{\end{picture}}

\newcount\rowcount
\newcount\columncount

\def\begintableau#1#2{                                                                  
\global\rowcount=#1
\begin{picture}(#2,#1)(0,0)
\linethickness{0.1pt}
}

\def\endtableau{\end{picture}}

\def\row#1{
        \global\advance\rowcount by -1
        \global\columncount=0
        \dorowlist#1\endList
}

\def\dorowlist{\afterassignment\handlenextentry\let\next=}

\def\handlenextentry{
        \ifx\next\endList
                \let\next=\relax
        \else
                \ifx\next o
                        \put(\columncount,\rowcount){\framebox(1,1){\space}}
                        \global\advance\columncount by 1
                \else
                        \ifx\next-
                                \global\advance\columncount by 1
                        \else
                                \put(\columncount,\rowcount){\framebox(1,1){$\next$}}
                                \global\advance\columncount by 1
                        \fi
                \fi
                \let\next=\dorowlist
        \fi
        \next
}

\newcount\ca
\newcount\cb

\def\doprintrowlist{\afterassignment\handlenextprintentry\let\next=}

\def\handlenextprintentry{
        \ifx\next\endList
                \let\next=\relax
        \else
                \put(\ca,\cb){\makebox(1,1){$\scriptstyle\next$}}
                \global\advance\ca by 1
                \let\next=\doprintrowlist
        \fi
        \next
}

\newtheorem{Theorem}{Theorem}
\newtheorem{Corollary}[Theorem]{Corollary}

\theoremstyle{remark}
\newtheorem*{Remark}{Remark}

\numberwithin{equation}{section}

\def\v#1{\left\vert #1\right\vert}
\def\eqref#1{(\ref{#1})}
\def\so{\operatorname{\text {\it so}}}
\def\sp{\operatorname{\text {\it sp}}}

\def\fl#1{\left\lfloor#1\right\rfloor}

\def\v#1{\left\vert #1\right\vert}
\def\eqref#1{(\ref{#1})}
\def\so{\operatorname{\text {\it so}}}
\def\sp{\operatorname{\text {\it sp}}}
\def\oddrows{\operatorname{oddrows}}
\def\poq#1#2{(#1;q)_#2}

\begin{document}
\title{Vicious walkers, friendly walkers and Young tableaux III: Between two walls}
\author{Christian Krattenthaler\thanks{ Research partially supported by EC's IHRP Programme,
grant HPRN-CT-2001-00272.}\\
 Institut f\"{u}r Mathematik, Universit\"{a}t Wien\\
Strudlhofgasse 4, A-1090 Vienna, Austria,\\
Anthony J. Guttmann
\thanks{ Research supported by the Australian Research Council.}
\thanks{ Author for communication. Phone (03)-8344-5550, fax (03)-8344-4599, tonyg@ms.unimelb.edu.au.}\\
 Department of Mathematics and Statistics, The
University of Melbourne\\ Victoria 3010, Australia\\
and Xavier G. Viennot$^*$\\
LaBRI, Universit\'{e} Bordeaux 1, 351 cours de la
Lib\'{e}ration,\\ 33405 Talence Cedex, France.\\\\
Running Head: Vicious walkers between two walls.}

\maketitle
\newpage

\begin{abstract}
We derive exact and asymptotic results for the number of star
and watermelon configurations of vicious walkers
confined to lie between two impenetrable walls, as well as corresponding
results for the analogous
problem of $\infty$-friendly walkers. Our proofs make use of
results from
symmetric function theory and the theory of basic hypergeometric series.\\
{\bf Keywords;} Vicious walkers, friendly walkers, symmetric functions,
affine Weyl groups.
\end{abstract}

\section{Introduction}
This is the third paper in a series studying vicious and friendly
walkers.
In  the first paper~\cite{GuOVAA} it was shown how certain
results from the theory of Young tableaux, and related
results in algebraic combinatorics enabled one to readily
prove closed form expressions for the number of
star and watermelon  configurations of vicious walkers on a $d$-dimensional
lattice.

In the second paper~\cite{GuKVAA}, we showed how some results from
the theory of symmetric functions could be used to prove analogous results
for the more difficult problem of walkers in the presence of
an impenetrable wall. We also gave rigorous asymptotic results.

In that paper we also developed the theory of $n$-friendly walkers,
introduced in~\cite{GV} and~\cite{Katori}. The two models differ slightly.
In \cite{GV}, the ``vicious''
constraint is systematically relaxed, so that any two walks
(but not more than two) may stay together for
up to $n$ lattice sites in a row, but may never swap sides. We refer
to this as the $n$-friendly walker model. In the limit as
$n \rightarrow \infty$ we obtain the $\infty$-friendly walker model
in which
two walkers may share an
arbitrary number of steps.
The Tsuchiya-Katori model \cite{Katori}, by contrast, corresponds to a
variant of the $\infty$-friendly walker model which allows
{\it any number}
of walkers to share any number of
lattice sites, whereas in the Guttmann-V\"oge definition \cite{GV},
only two walkers may share a
lattice site. We subsequently refer to these
two models as the TK and GV models respectively. Thus the number of TK
friendly walk configurations
gives an upper bound on the number of $\infty$-friendly walk configurations
in the definition of GV. We make use of this
observation in subsequent proofs.

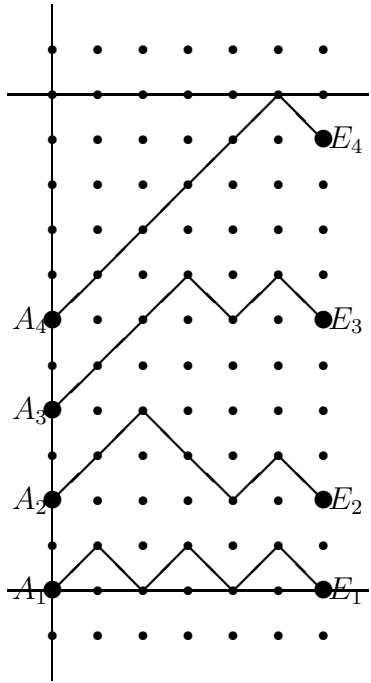
\begin{figure}[t]
$$
        \Gitter(7,13)(0,-1)
        \Koordinatenachsen(7,13)(0,-1)
        \PfadDicke{.5pt}
        \Pfad(-1,11),11111111\endPfad
        \PfadDicke{1pt}
        \Pfad(0,0),343434\endPfad
        \Pfad(0,2),334434\endPfad
        \Pfad(0,4),333434\endPfad
        \Pfad(0,6),333334\endPfad
        \DickPunkt(0,0)
        \DickPunkt(0,2)
        \DickPunkt(0,4)
        \DickPunkt(0,6)
        \DickPunkt(6,0)
        \DickPunkt(6,2)
        \DickPunkt(6,6)
        \DickPunkt(6,10)
        \Label\l{A_1}(0,0)
        \Label\l{A_2}(0,2)
        \Label\l{A_3}(0,4)
        \Label\l{A_4}(0,6)
        \Label\r{E_1}(6,0)
        \Label\r{E_2}(6,2)
        \Label\r{E_3}(6,6)
        \Label\r{E_4}(6,10)
\begin{picture}(15,5)(0,0)
\end{picture}
$$
\vskip.3cm
\caption{A star of $p=4$ vicious walkers, of length $m=6$, confined
to a strip of width $h=11.$}
\label{F1}
\end{figure}

In this, the final paper in the series, we address the problem of
vicious and friendly walkers confined to a finite strip --- or, equivalently,
confined to lie between two parallel walls. A star configuration in a strip
of width 11 is shown in Figure~\ref{F1}.

{\em Vicious walkers}
describes the situation in which two or more walkers arriving
at the same lattice site annihilate one another. Accordingly,
the only configurations we consider in that case
are those in which such contacts
are forbidden. Alternatively expressed, we consider mutually
self-avoiding networks of lattice walks which also model
directed polymer
networks. The connection of these vicious walker problems to the
5 and  6~vertex model of statistical mechanics was also discussed
in \cite{GuOVAA}.

The problem, together with a number of physical applications,
was first introduced by Fisher \cite{FIS}. Physical applications
include models of wetting, and Fisher's original articles already
raised the physical interesting consequences of the introduction
of geometrical constraints in the form of walls.
Very recently, it was shown in \cite{DL02} that the problem of vicious walkers
in a
(periodic) strip involves precisely the same combinatorics as arises in
three-dimensional Lorentzian quantum gravity.
The general model is one of $p$ random walkers on a
$d$-dimensional lattice who at
regular time intervals simultaneously take one step with equal
probability in the direction of one of the allowed lattice vectors
such that at no time do two walkers occupy the same lattice site.

Very recently, a number of authors \cite{Joh00,Pra100,Pra200,Ba00} have
made fascinating connections between certain properties of two-dimensional
vicious walkers and the eigenvalue distribution of
certain random
matrix ensembles. In \cite{Joh00} a model is introduced
which can be considered as a randomly growing Young diagram,
or a totally asymmetric one-dimensional exclusion process.
(This could be interpreted in the vicious walker model where at each
time unit {\em exactly one} of the walkers moves. This model
occurs already in \cite{FIS}.)
It is shown that the appropriately scaled shape fluctuations converge
in distribution to the Tracy-Widom distribution \cite{TWGUE}
of the largest
eigenvalue of the Gaussian Unitary Ensemble (GUE). Similarly, in
\cite{Ba00} a vicious
walker model is considered in which the end-point
fluctuations of the top-most walker (in our notation) are considered.
In that case the appropriately scaled limiting distribution is
that of the largest eigenvalue of another distribution, the Gaussian
Orthogonal Ensemble (GOE) \cite{TWGOE}. Finally in \cite{Pra100,Pra200}
 the height
distribution of a given point in the substrate of a one-dimensional
growth process is considered, and this is generalised to models in
the Kardar-Parisi-Zhang (KPZ) universality class \cite{Ka86}.
The configurations considered again appear like vicious walkers. Again
fluctuations and other properties of the models are found that follow
GOE or GUE distributions.

The two standard topologies of interest are that of a \textit{star}
and a \textit{watermelon}. Consider a directed square lattice,
rotated
$45^\circ$ and augmented by a factor of $\sqrt2$,
so that the ``unit" vectors on the lattice are
$(1,1)$ and
$(1,-1)$.
Both configurations consist of $p$
branches of length $m$ (the lattice paths along which the walkers proceed)
which start at $(0,0),(0,2),(0,4),\dots,(0,2p-2).$
The watermelon configurations end at
$(m,k),(m,2+k),(m,4+k),\dots,(m,k+2p-2)$,
for some $k$. For stars, the end points
of the
branches all have $x$-coordinate equal to $m,$ but the $y$
 coordinates
are unconstrained, apart from the ordering imposed by the
non-touching
condition. Thus if the end points are
$(m,e_1),(m,e_2),(m,e_3),\cdots,(m,e_p),$ then
$e_1 < e_2 < e_3 < \cdots < e_p \le 2p-2+m.$ In the
problem considered
here, the additional constraint of impenetrable walls
imposes the conditions that at no stage
may any walker step to a point with negative $y$-coordinate, or
to a point with $y$-cordinate greater than the strip width, $h$.
We can also consider {\em displaced configurations,} in which
the starting points are $(0,a),(0,a+2),(0,a+4),\dots,(0,a+2p-2).$

Vicious walkers confined to lie between two walls
can be alternatively viewed as random
walks in an alcove of an affine Weyl group of type $C$ if the
set of allowed steps is appropriately chosen.
In this form they are considered by Grabiner in \cite[Sec.~5]{Gr01}.
The topic of the paper \cite{Gr01} is the exact enumeration of
random walks in alcoves of affine Weyl groups of types $A$, $C$ and
$D$.
One of the problems posed in \cite{Gr01} is to find asymptotic
formulae for these random walks (when the length of the walks
goes to infinity). We solve this problem for the random
walks in an alcove of type $C$
that correspond to our vicious walkers between two walls.
Asymptotic results for the other enumeration problems considered in
\cite{Gr01} will appear in the forthcoming paper \cite{KratBT}.
An earlier, related paper \cite{PF90}
contains partial results for vicious walkers on a cylinder
in the case of an odd number of walkers (which are equivalent to
random walkers in an alcove of type $A$).

It is intuitively clear that the asymptotic growth of the number of
vicious walkers
within a strip must be exponential, with the base of the
exponential depending on {\em both} the width of the strip and the number of
walkers, but not on the starting and end points of the walkers.
This is confirmed by our results. (The same must of course
be true for $n$-friendly walkers, although we are only able to
rigorously confirm this for $\infty$-friendly walkers in the TK model, by
deriving explicit formulae.)
Thus, for example, the asymptotics of stars and watermelons in the
same strip will be exponential with the same base, and will only
differ in the constant by which the exponential is
multiplied. In the cases of walks with only one
wall, or no walls \cite{GuOVAA,GuKVAA}, the asymptotic growth factor
is just $2^p,$ where $p$ is the number of walkers.

Our results explicitly establish these intuitive results. Furthermore,
in proving these results we present a variety of mathematical
techniques and results which are likely to be of value in the
study of related problems in the mathematical physics
literature, a number of which are discussed above.

Our paper is organised as follows.
In Section~2 we provide exact formulas for the number of vicious
walkers between two walls with arbitrary starting and end points.
With the exception of one, these appear already in \cite{Gr01},
in equivalent forms. These results follow
from the Lindstr\"om--Gessel--Viennot
theorem on nonintersecting lattice paths and known results for
lattice paths between two parallel lines.
They express the number of vicious walkers within
a strip as determinants. In Section~3 we address the asymptotics of
these formulas when the length of the branches of the vicious walkers
goes to infinity. In Theorem~\ref{thm:vicious-with-2-asy} we give the
asymptotics for vicious walkers within a strip for arbitrary (but
fixed) starting and end points. By specializing the starting and
end points, we obtain asymptotic results for watermelons within a
strip, see Corollary~\ref{thm:watermelons-with-2-asy}. In order to
obtain asymptotic results for stars within a strip, the formula in
Theorem~\ref{thm:vicious-with-2-asy} has to be summed over all possible
end points. To carry out this summation is a highly nontrivial task. It
requires some symmetric function theory, in particular certain
relations between Schur functions and symplectic and orthogonal
characters, and a summation theorem for a basic hypergeometric
series. The final result is given in
Theorem~\ref{thm:starvicious-with-2}. This theorem is then
specialized to obtain the asymptotics for stars within a strip, see
Corollaries~\ref{thm:star-with-2} and \ref{thm:infty-star-with-2}.

\section{The number of vicious walker configurations with arbitrary fixed
starting and end points}

The
Lindstr\"om--Gessel--Viennot determinant~\cite{LindAA,GeViAB}
in the case of the presence of two
walls yields the following result.
It appears already in
\cite[(13)]{Gr01}, in an equivalent form.

\begin{Theorem} \label{thm:vicious-with-2}
Let $0\le a_1<a_2<\dots<a_p\le h$, all $a_i$'s of the same parity, and
$0\le e_1<e_2<\dots<e_p\le h$, all $e_i$'s of the same parity, such that
$a_i+e_i\equiv m\pmod2$, $i=1,2,\dots,p$.
The number of vicious walkers with $p$ branches of length $m$, the
$i$-th branch running from $A_i=(0,a_i)$ to $E_i=(m,e_i)$,
$i=1,2,\dots,p$,
which do not go below the $x$-axis nor above the line $y=h$,
is given by
\begin{equation} \label{eq:vicious-with-2}
\det_{1\le s,t\le p}\(\sum _{k=-\infty} ^{\infty}
\(\binom {m} {\frac {m+e_t-a_s}
{2}+k(h+2)}-\binom {m} {\frac {m+e_t+a_s} {2}+k(h+2)+1}\)\).
\end{equation}
\end{Theorem}

\begin{proof}According to the main theorem of
non-intersecting lattice paths
\cite[Cor.~2]{GeViAB} (see \cite[Theorem~1.2]{StemAE}), the number of
vicious walkers in
question equals
\begin{equation} \label{e2}
\det_{1\le s,t\le p}\big(\v{\mathcal P^{++}\big(A_t\to E_s\big)}\big),
\end{equation}
where $\mathcal P^{++}\big(A\to E\big)$ denotes the set of all lattice paths from $A$ to
$E$ which do not go below the $x$-axis nor above the line $y=h$.
There is a well-known formula (see \cite[(1.7)]{MohaAE}) for the latter
number, which is obtained by an iterated reflection principle.
Substitution of this formula into \eqref{e2} immediately
gives \eqref{eq:vicious-with-2}.
\end{proof}

By bringing the sums in \eqref{eq:vicious-with-2} outside the
determinant (using the multi-linearity of the determinant), the number
of these vicious walkers can be described as a multiple sum of
determinants. In some cases, such as for certain watermelons and stars,
the determinants can be evaluated. In those cases, a multiple hypergeometric
sum is obtained.

\begin{Theorem} \label{thm:vicious-with-2a}
Let $0\le e_1<e_2<\dots<e_p\le h$ with $e_i\equiv m\pmod2$, $i=1,2,\dots,p$.
The number of vicious walkers with $p$ branches of length $m$, the
$i$-th branch running from $A_i=(0,2i-2)$ to $E_i=(m,e_i)$,
which do not go below the $x$-axis nor above the line $y=h$,
$i=1,2,\dots,p$,
is given by
\begin{multline} \label{eq:vicious-with-2a}
\sum _{k_1,\dots,k_p=-\infty} ^{\infty}
2^{p-p^2}\prod _{s=1} ^{p}\frac
{(e_s+2k_s(h+2)+1)\,(m+2s-2)!} {\(\frac {m+e_s}
{2}+k_s(h+2)+p\)!\,\(\frac {m-e_s} {2}-k_s(h+2)+p-1\)!}\\
\times
\prod _{1\le s<t\le p} ^{}(e_t-e_s+2(h+2)(k_t-k_s))
(e_t+e_s+2(h+2)(k_t+k_s)+2).
\end{multline}
\end{Theorem}
\begin{proof}As described above the statement of the theorem, we
first write the expression \eqref{eq:vicious-with-2}, with $a_i=2i-2$, as a
sum of determinants,
\begin{equation} \label{eq:sum-of-dets}
\sum _{k_1,\dots,k_p=-\infty} ^{\infty}\det_{1\le s,t\le p}
\(\binom {m} {\frac {m+e_t}
{2}+k_t(h+2)-s+1}-\binom {m} {\frac {m+e_t} {2}+k_t(h+2)+s}\).
\end{equation}
Suppose that, initially, we disregard the terms $k_t(h+2)$ in the determinant,
then it simplifies to
\begin{equation} \label{eq:sum-of-dets2}
\det_{1\le s,t\le p}
\(\binom {m} {\frac {m+e_t}
{2}-s+1}-\binom {m} {\frac {m+e_t} {2}+s}\).
\end{equation}
Gessel--Viennot theory (again) says that this determinant counts
vicious walkers with $p$ branches of length $m$, the
$i$-th branch running from $A_i=(0,2i-2)$ to $E_i=(m,e_i)$,
which do not go below the $x$-axis. By Theorem~6 of
\cite{GuKVAA}, the number of these vicious walkers is given by
\begin{equation} \label{e20}
2^{-p^2+p}\prod _{s=1} ^{p}\frac {(e_s+1)\,(m+2s-2)!}
{\(\frac {m+e_s} {2}+p\)!\,\(\frac {m-e_s} {2}+p-1\)!}
\prod _{1\le s<t\le p}
^{}(e_t-e_s)(e_s+e_t+2).
\end{equation}
Hence, the determinant in \eqref{eq:sum-of-dets2} must equal the
expression \eqref{e20}. In fact, as the equality between
\eqref{eq:sum-of-dets2} and \eqref{e20} can be reduced to an equation
which is polynomial in $e_1,e_2\dots,e_p$, the equality is true for
{\em any} choice of $e_1,e_2,\dots,e_p$. In particular, it remains
true if we replace $e_i$ by $e_i+2k_i(h+2)$, $i=1,2,\dots,p$.
However, the
determinant in \eqref{eq:sum-of-dets2} under these replacements becomes
the determinant in \eqref{eq:sum-of-dets}. Thus, if we substitute the
expression \eqref{e20}, with these replacements, into
\eqref{eq:sum-of-dets}, we immediately obtain \eqref{eq:vicious-with-2a}.
\end{proof}

For the subsequent asymptotic calculations, however, we need a
different type of expression for the number of vicious walkers under
consideration.
This expression can be easily derived by a combination of the
Lindstr\"om--Gessel--Viennot theorem and an alternative expression for
the number of lattice paths between two parallel boundaries in terms
of sines and cosines.
It appears already in
\cite[(18)]{Gr01}, in an equivalent form.

\begin{Theorem} \label{thm:vicious-with-2b}
Let $0\le a_1<a_2<\dots<a_p\le h$, all $a_i$'s of the same parity, and
$0\le e_1<e_2<\dots<e_p\le h$, all $e_i$'s of the same parity, such that
$a_i+e_i\equiv m\pmod2$, $i=1,2,\dots,p$.
The number of vicious walkers with $p$ branches of length $m$, the
$i$-th branch running from $A_i=(0,a_i)$ to $E_i=(m,e_i)$,
$i=1,2,\dots,p$,
which do not go below the $x$-axis and not above the line $y=h$,
is given by
\begin{equation} \label{eq:vicious-with-2b}
\frac {2^p} {(h+2)^p}
\sum _{k_1,\dots,k_p=1} ^{h+1}
\bigg(2^p \prod _{s=1}
^{p}\cos\frac {k_s\pi} {h+2}\bigg)^m
\prod _{t=1} ^{p}\sin\frac {\pi k_t(e_t+1)} {h+2}
\det_{1\le s,t\le p}
\(\sin\frac {\pi k_t(a_s+1)} {h+2}\).
\end{equation}
\end{Theorem}

\begin{proof}We already know that, by the
Lindstr\"om--Gessel--Viennot theorem, the number in question is
given by \eqref{e2}. Instead of using the iterated reflection formula
for
$\v{\mathcal P^{++}\big(A\to E\big)}$, we now apply the (equally
well-known) alternative
formula (see \cite[\S184, Ex.~1, Eq.~(9)]{JordAA})
$$\v{\mathcal P^{++}\big(A\to E\big)}=\frac {2} {h+2}\sum _{k=1} ^{h+1}
\(2\cos \frac {\pi k} {h+2}\)^m\sin\frac {\pi k(a+1)} {h+2}\cdot \sin\frac
{\pi k(e+1)} {h+2},$$
given that $A=(0,a)$ and $E=(m,e)$.
Substituting this
into \eqref{e2}, and bringing the summations and a few
factors outside
of the determinant utilising the multi-linearity of the determinant,
we get \eqref{eq:vicious-with-2b}.
\end{proof}

\section{The asymptotics of vicious walkers between two walls}

We shall now use the exact formulae from the previous section to
derive asymptotic formulae for vicious walkers between two walls.
In the first subsection, we address the case where the end points
of the vicious walkers are kept fixed. By summing over all possible
end points, we shall then obtain asymptotic formulae for stars
in the second subsection.

\subsection{Vicious walkers with fixed end points}

Theorem~\ref{thm:vicious-with-2b} enables us to derive asymptotic
formulae for the number of vicious walkers between two walls, for
arbitrary starting and end points.

\begin{Theorem} \label{thm:vicious-with-2-asy}
Let $0\le a_1<a_2<\dots<a_p\le h$, all $a_i$'s of the same parity, and
$0\le e_1<e_2<\dots<e_p\le h$, all $e_i$'s of the same parity, such that $a_i+e_i\equiv m\pmod2$,
$i=1,2,\dots,p$.
The number of vicious walkers with $p$ branches of length $m$, the
$i$-th branch running from $A_i=(0,a_i)$ to $E_i=(m,e_i)$,
$i=1,2,\dots,p$,
which do not go below the $x$-axis nor above the line $y=h$,
is asymptotically
\begin{multline} \label{eq:vicious-with-2-asy}
 \frac {4^{p^2}} {(h+2)^p} \bigg(2^p \prod _{s=1}
^{p}\cos\frac {s\pi} {h+2}\bigg)^m
 \prod _{1\le s<t\le p} ^{}\sin\frac {\pi (a_t-a_s)} {2(h+2)} \cdot
   \sin\frac {\pi (e_t-e_s)} {2(h+2)}\\
\times
 \prod _{1\le s\le t\le p}\sin\frac {\pi (a_t+a_s+2)} {2(h+2)} \cdot
   \sin\frac {\pi (e_t+e_s+2)} {2(h+2)}.
\end{multline}
\end{Theorem}

\begin{Remark}What this theorem says is that vicious walkers in a
strip of width $h$ grow exponentially like $\big(2^p \prod _{s=1}
^{p}\cos \frac {s\pi} {h+2}\big)^m$, everything else is just the
multiplicative constant.
In particular, specialising either to watermelons or stars leads to
the same dominant asymptotic behaviour, with only a multiplicative constant
changing as the configurations change.
\end{Remark}

\begin{proof}Theorem~\ref{thm:vicious-with-2b} tells us that the
number of vicious walkers that we wish to estimate can be written in
the form of a finite sum $\sum _{\ell} ^{}c_\ell b_\ell^m$, where the
$c_\ell$'s and $b_\ell$'s are independent of $m$. Hence,
what we have to find is $b=\max_\ell b_\ell$. Then, asymptotically,
the number of vicious walkers is $b^m\sum _{\ell:\ b_\ell=b}
^{}c_\ell$.

The $b_\ell$'s have the form
\begin{equation} \label{eq:b_l}
2^p \prod _{s=1}
^{p}\cos \big({k_s\pi} / (h+2)\big).
\end{equation}
To find the largest among these, we have
to choose $k_s$ close to the lower limit of the summation in
\eqref{eq:vicious-with-2b}, which is $1,$ or
close to the upper limit, $h+1$. However, we are not allowed to
make a choice such that $k_s=k_t$ for some $s\ne t$, because in that
case the determinant in \eqref{eq:vicious-with-2b} vanishes. For the same
reason a choice such that $k_s=h+2-k_t$ for some $s$ and $t$ is forbidden.
Therefore the expression \eqref{eq:b_l} will be maximal if the set
$\{k_1,k_2\dots,k_p\}$ is chosen from
$\{1,2,\dots,p,h+2-p,\dots,h,h+1\}$, subject to the two restrictions
mentioned above. These conditions give exactly $2^p$ different choices. As a
short calculation shows, the sum of the corresponding summands in
\eqref{eq:vicious-with-2b} equals
\begin{equation*}
 \frac {4^p} {(h+2)^p} \bigg(2^p \prod _{s=1}
^{p}\cos\frac {s\pi} {h+2}\bigg)^m
\det_{1\le s,t\le p}
\( \sin\frac {\pi t(a_s+1)} {h+2}\)
\det_{1\le s,t\le p}
\( \sin\frac {\pi t(e_s+1)} {h+2}\),
\end{equation*}
or, equivalently,
\begin{multline} \label{eq:prod-det}
 \frac {(-1)^p} {(h+2)^p} \bigg(2^p \prod _{s=1}
^{p}\cos\frac {s\pi} {h+2}\bigg)^m
\det_{1\le s,t\le p}
\( e^{\frac {\pi it} {h+2}(a_s+1)}-
e^{-\frac {\pi it} {h+2}(a_s+1)}\)\\
\times
\det_{1\le s,t\le p}
\( e^{\frac {\pi it} {h+2}(e_s+1)}-
e^{-\frac {\pi it} {h+2}(e_s+1)}\).
\end{multline}
Both determinants are easily evaluated by means of the determinant identity
\begin{align} \label{eq:sympl}
\det\limits_{1\le i,j\le N}\(x_i^{j}-x_i^{-j}\)&=
(x_1x_2\cdots x_N)^{-N}
\prod _{1\le i<j\le N} ^{}(x_i-x_j)(1-x_ix_j)\prod _{i=1}
^{N}(x_i^2-1),
\end{align}
which may be readily proved by the standard argument that
proves Vandermonde-type determinant identities.

A little manipulation then leads to \eqref{eq:vicious-with-2-asy}.
\end{proof}

If we now specialize $a_i$ to $a+2i-2$ and $e_i$ to $e+2i-2$ in
Theorem~\ref{thm:vicious-with-2-asy}, we then obtain the asymptotics
for watermelons between two walls.

\begin{Corollary} \label{thm:watermelons-with-2-asy}
Let $a$ and $e$ be integers with $0\le a,e\le h-2p+2$ and $a+e\equiv
m\pmod2$.
The number of watermelons with $p$ branches of length $m$, in which
the lowest branch starts at height $a$ and terminates at height $e$,
which do not go below the $x$-axis nor above the line $y=h$,
is asymptotically
\begin{multline} \label{eq:watermelons-with-2-asy}
 \frac {4^{p^2}} {(h+2)^p} \bigg(2^p \prod _{s=1}
^{p}\cos\frac {s\pi} {h+2}\bigg)^m
 \prod _{1\le s<t\le p} ^{}\sin^2\frac {\pi (t-s)} {(h+2)} \\
\times
 \prod _{1\le s\le t\le p}\sin\frac {\pi (a+t+s-1)} {(h+2)} \cdot
   \sin\frac {\pi (e+t+s-1)} {(h+2)}.
\end{multline}

Let $a$ and $e$ be integers with $0\le a,e\le h-2p+2$ and $a+e\equiv
m\pmod2$.
The number of $\infty$-friendly
watermelons
in the TK model with $p$ branches of length $m$, in which
the lowest branch starts at height $a$ and terminates at height $e$,
which do not go below the $x$-axis nor above the line $y=h$,
is asymptotically
\begin{multline} \label{eq:infty-watermelons-with-2-asy}
 \frac {4^{p^2}} {(h+2p)^p} \bigg(2^p \prod _{s=1}
^{p}\cos\frac {s\pi} {h+2p}\bigg)^m
 \prod _{1\le s<t\le p} ^{}\sin^2\frac {2\pi (t-s)} {(h+2p)} \\
\times
 \prod _{1\le s\le t\le p}\sin\frac {\pi (a+2t+2s-3)} {(h+2p)} \cdot
   \sin\frac {\pi (e+2t+2s-3)} {(h+2p)}.
\end{multline}
\end{Corollary}

\begin{proof}There is nothing to say about the first claim, which
follows immediately from the theorem.
To establish the second claim, we shift the $i$-th branch of the
$\infty$-friendly watermelon
by $2(i-1)$ units up,
as in the proof of Theorem~4 of \cite{GuKVAA}.
We transform $\infty$-friendly watermelons into families of
lattice paths which do not touch one another by shifting the $i$-th path up by
$2(i-1)$ units.
Thus we obtain a set of vicious walkers
with $p$ branches of length $m$, the
$i$-th branch starting from $A_i=(0,a+4i-4)$ and terminating at
$E_i=(m,e+4i-4)$, $i=1,2,\dots,p$,
which do not go below the $x$-axis and not above the line $y=h+2p-2$~(!).
Hence, Theorem~\ref{thm:vicious-with-2-asy} with $a_i=a+4i-4$, $e_i=e+4i-4$,
$i=1,2,\dots,p$, and $h$ replaced by $h+2p-2$ immediately gives the
desired asymptotics.
\end{proof}

Clearly, by performing the obvious summations of
\eqref{eq:watermelons-with-2-asy}
over $e$, respectively
$a$, we could also obtain
the asymptotics for watermelons of arbitrary deviation. The resulting
sums do not appear to simplify however.
Nevertheless, since the summations are over finite sets (depending
only on the width $h$ of the strip and the number $p$ of walkers),
it is obvious that the order of
the asymptotic growth is again $\big(2^p \prod _{s=1}
^{p}\cos \frac {s\pi} {h+2}\big)^m$ for vicious walkers and
$\big(2^p \prod _{s=1}
^{p}\cos \frac {s\pi} {h+2p}\big)^m$ for $\infty$-friendly walkers in
the TK model.

It should be noted, however, that in contrast to watermelons
without restriction and
with the restriction of {\em one} wall, as considered in our previous
papers, the situation considered here, that is in the presence
of the restriction of
{\em two} walls, the asymptotics of (ordinary)
watermelons and $\infty$-friendly watermelons (compare the bases of the
exponentials in the two
statements in Corollary~\ref{thm:watermelons-with-2-asy})
is of a different order of
magnitude (except in the case of single branch watermelons, of course).
Hence, from these considerations, it is impossible to
conclude whether $n$-friendly watermelons restricted by two walls will have
the same order of magnitude as (ordinary) watermelons, or not.

\subsection{Asymptotics for stars}

Let us turn now to the asymptotics for stars. We begin with a
general theorem, which solves the problem, posed in \cite{Gr01}, of
computing the asymptotics for the number of
random walks in an alcove of an affine Weyl group of type $C$ if the
allowed steps are of the form $\frac {1} {2}(\pm1,\pm1,\dots,\pm1)$.
\begin{Theorem} \label{thm:starvicious-with-2}
Let $0\le a_1<a_2<\dots<a_p\le h$ be integers, all of the same parity.
The number of vicious walkers with $p$ branches of length $m$, the
$i$-th branch starting from $A_i=(0,a_i)$,
$i=1,2,\dots,p$,
which do not go below the $x$-axis nor above the line $y=h$,
is asymptotically
\begin{multline} \label{eq:starvicious-with-2even}
\frac {4^{p^2}} {(h+2)^p}  \bigg(2^p \prod _{s=1} ^{p}\cos \frac {s \pi}
{h+2}\bigg)^m
 \prod _{1\le s<t\le p} ^{}\sin \frac {\pi  (a_t-a_s)} {2(h+2)} \cdot
    \sin \frac {\pi  (t-s)} {h+2}\\
\times
 \prod _{1\le s\le t\le p} ^{}\sin \frac {\pi  (a_t+a_s+2)} {2(h+2)} \cdot
  \sin \frac {\pi  (t+s)} {h+2}
 \prod _{s=0} ^{p}\prod _{t=1} ^{p}\frac {\sin \frac
   {\pi  (t-s+\fl{(h+2)/2})}
   {h+2}} {\sin \frac {\pi  (t-s+p)} {h+2}}\\
\times
 \prod _{s=1} ^{p}\frac {\sin \frac {(s+\fl{(h+2)/2}-p) \pi} {h+2}}
   {\sin \frac {(2s+\fl{(h+2)/2}-p) \pi} {h+2}},
\end{multline}
if $m+a_i$ is even, and
\begin{multline} \label{eq:starvicious-with-2odd}
\frac {4^{p^2}} {(h+2)^p}  \bigg(2^p \prod _{s=1} ^{p}\cos \frac {s \pi}
{h+2}\bigg)^m
 \prod _{1\le s<t\le p} ^{}\sin \frac {\pi  (a_t-a_s)} {2(h+2)} \cdot
    \sin \frac {\pi  (t-s)} {h+2}\\
\times
 \prod _{1\le s\le t\le p} ^{}\sin \frac {\pi  (a_t+a_s+2)} {2(h+2)}
\cdot   \sin \frac {\pi  (t+s)} {h+2}
 \prod _{s=0} ^{p}\prod _{t=1} ^{p}\frac {\sin \frac
   {\pi  (t-s+\fl{(h+1)/2})}
   {h+2}} {\sin \frac {\pi  (t-s+p)} {h+2}},
\end{multline}
if $m+a_i$ is odd.
\end{Theorem}

\begin{proof}It is obvious that, in view of
Theorem~\ref{thm:vicious-with-2-asy}, we have to compute the sum of
\eqref{eq:vicious-with-2-asy} over all possible choices of
$e_1<e_2<\dots<e_p$. Here we have to distinguish between two cases,
depending on whether $m+a_i$ is even or odd.

First let $m+a_i$ be odd. This implies that all the $e_i$'s are odd
as well, so that we have to compute the sum of
\eqref{eq:vicious-with-2-asy} over all possible choices of
$1\le e_1<e_2<\dots<e_p\le h$, with  $e_i=2e'_i-1$ for some integer $e'_i$,
$i=1,2,\dots,p$. If we remember that
expression \eqref{eq:vicious-with-2-asy} came from
\eqref{eq:prod-det}, we see that this is
\begin{multline} \label{eq:vicious-with-2-asy-sum}
 \frac {2^{p^2}} {(h+2)^p} \bigg(2^p \prod _{s=1}
^{p}\cos\frac {s\pi} {h+2}\bigg)^m
 \prod _{1\le s<t\le p} ^{}\sin\frac {\pi (a_t-a_s)} {2(h+2)}
 \prod _{1\le s\le t\le p}\sin\frac {\pi (a_t+a_s+2)} {2(h+2)} \\
\times (-i)^p
\sum _{1\le e'_1<\dots<e'_p\le \fl{(h+1)/2}} ^{}
\det_{1\le s,t\le p}
\( e^{\frac {2\pi it} {h+2}e'_s}-
e^{-\frac {2\pi it} {h+2}e'_s}\).
\end{multline}
In order to evaluate the sum in the last line, we rewrite it in terms
of symplectic characters
$\sp_\la(x_1^{\pm1},x_2^{\pm1},\dots,x_p^{\pm1})$,
which are
defined by (see \cite[(24.18)]{FuHaAA})
\begin{equation} \label{e24}
\sp_\la(x_1^{\pm1},x_2^{\pm1},\dots,x_p^{\pm1})
=\frac {\det\limits_{1\le s,t\le
p}(x_t^{\la_s+p-s+1}-x_t^{-(\la_s+p-s+1)})}
{\det\limits_{1\le s,t\le
p}(x_t^{p-s+1}-x_t^{-(p-s+1)})}.
\end{equation}
Therefore, writing $q$ for $e^{2\pi i/(h+2)}$ and $H$ for
$\fl{(h+1)/2}$, the sum in the last
line of \eqref{eq:vicious-with-2-asy-sum} equals
\begin{multline} \label{eq:sum-sympl}
\det_{1\le s,t\le p}
\( q^{st}-q^{-st}\)
\sum _{1\le e'_1<\dots<e'_p\le H} ^{}
\sp_{(e'_p-p,\dots,e'_1-1)}(q^{\pm1},q^{\pm2},\dots,q^{\pm p})\\
=\det_{1\le s,t\le p}
\( q^{st}-q^{-st}\)
\sum _{\nu\subseteq ((H-p)^p)} ^{}
\sp_\nu(q^{\pm1},q^{\pm2},\dots,q^{\pm p}).
\end{multline}
Now we appeal to the formula (see
\cite[(3.4)]{KratBC}),
\begin{equation} \label{e31}
s_{(c^r)}(x_1,x_1^{-1},\dots,x_p,x_p^{-1},1)=
\sum _{\nu\subseteq (c^r)} ^{}\sp_\nu(x_1^{\pm1},\dots,x_p^{\pm1}).
\end{equation}
Use of this formula in \eqref{eq:sum-sympl} gives
\begin{equation} \label{eq:sum-schur}
\det_{1\le s,t\le p}
\( q^{st}-q^{-st}\)
s_{((H-p)^p)}(q^{-p},\dots,q^{-1},1,q,\dots,q^p).
\end{equation}
Clearly, the determinant is easily evaluated by means of
\eqref{eq:sympl}, whereas the specialized Schur function can be
evaluated by means of the hook-content formula (see
\cite[I, Sec.~3,  Ex.~1]{MacdAC}, \cite[Ex.~A.30, (ii)]{FuHaAA})
\begin{equation} \label{e7}
s_\la(q^{-L},q^{-L+1},\dots,q^{-L+P})=q^{\sum _{l\ge1} ^{}(l-L-1)\la_l}
\prod _{\rh\in\la} ^{}\frac {1-q^{P+c_\rh}} {1-q^{h_\rh}},
\end{equation}
where $c_\rh$ and $h_\rh$ are the {\em content\/} and the {\em hook
length\/} of the cell $\rh$. Substitution of all this in
\eqref{eq:vicious-with-2-asy-sum} and some manipulation then leads to
\eqref{eq:starvicious-with-2odd}.

Now let $m+a_i$ be even. This implies that all the $e_i$'s are even
as well, so that we have to compute the sum of
\eqref{eq:vicious-with-2-asy} over all possible choices of
$0\le e_1<e_2<\dots<e_p\le h$, with $e_i=2e'_i-2$ for some integer $e'_i$,
$i=1,2,\dots,p$. Again, if we remember that
expression \eqref{eq:vicious-with-2-asy} came from
\eqref{eq:prod-det}, we see that this is
\begin{multline} \label{eq:vicious-with-2-asy-sum-even}
 \frac {2^{p^2}} {(h+2)^p} \bigg(2^p \prod _{s=1}
^{p}\cos\frac {s\pi} {h+2}\bigg)^m
 \prod _{1\le s<t\le p} ^{}\sin\frac {\pi (a_t-a_s)} {2(h+2)}
 \prod _{1\le s\le t\le p}\sin\frac {\pi (a_t+a_s+2)} {2(h+2)} \\
\times (-i)^p
\sum _{1\le e'_1<\dots<e'_p\le \fl{(h+2)/2}} ^{}
\det_{1\le s,t\le p}
\( e^{\frac {2\pi it} {h+2}(e'_s-\frac {1} {2})}-
e^{-\frac {2\pi it} {h+2}(e'_s-\frac {1} {2})}\).
\end{multline}
This time, it is possible to rewrite the sum in the last line in
terms of odd orthogonal characters.
The odd orthogonal characters $\so_\la
(x_1^{\pm1},x_2^{\pm1},\dots,x_m^{\pm1},1)$ where $x_1^{\pm1}$ is
a shorthand notation
for $x_1,x_1^{-1}$, etc., and $\la$ is an
$m$-tuple $(\la_1,\la_2,\dots,\la_m)$ of integers, or of
half-integers, are defined by
\begin{equation} \label{e12}
\so_\la(x_1^{\pm1},x_2^{\pm1},\dots,x_m^{\pm1},1)=\frac {\det\limits_{1\le i,j\le
m}(x_j^{\la_i+m-i+1/2}-x_j^{-(\la_i+m-i+1/2)})}
{\det\limits_{1\le i,j\le
m}(x_j^{m-i+1/2}-x_j^{-(m-i+1/2)})}.
\end{equation}
While Schur functions are polynomials in
$x_1,x_2,\dots,x_m$, odd orthogonal characters\break
$\so_\la(x_1^{\pm1},x_2^{\pm1},\dots,x_m^{\pm1},1)$ are polynomials
in $x_1,x_1^{-1},x_2,x_2^{-1},\dots,x_m,x_m^{-1},1$. They have
combinatorial descriptions in terms of certain tableaux as well, see
\cite[Sec.~2]{FuKrAA}, \cite[Sec.~6--8]{ProcAK}, \cite[Theorem~2.3]{SunaAE}.

Now, writing $q$ for $e^{2\pi i/(h+2)}$ and $H$ for
$\fl{(h+2)/2}$, the sum in the last
line of \eqref{eq:vicious-with-2-asy-sum-even} equals
\begin{multline} \label{eq:sum-ortho}
\det_{1\le s,t\le p}
\( q^{(s-\frac {1} {2})t}-q^{-(s-\frac {1} {2})t}\)
\sum _{1\le e'_1<\dots<e'_p\le H} ^{}
\so_{(e'_p-p,\dots,e'_1-1)}(q^{\pm1},q^{\pm2},\dots,q^{\pm p},1)\\
=\det_{1\le s,t\le p}
\( q^{(s-\frac {1} {2})t}-q^{-(s-\frac {1} {2})t}\)
\sum _{\nu\subseteq ((H-p)^p)} ^{}
\so_\nu(q^{\pm1},q^{\pm2},\dots,q^{\pm p},1).
\end{multline}
Again there is a formula which allows us to evaluate the sum in the last
line (see \cite[(3.2)]{KratBC}),
\begin{equation} \label{eq:Schur-orthosum}
s_{((c^{r-\ell},(c-1)^\ell)}(x_1,x_1^{-1},\dots,x_p,x_p^{-1},1)=\underset
{\oddrows\!\big((c^r)/\nu\big)=\ell}
{\sum _{\nu\subseteq (c^r)} ^{}}
\so_\nu(x_1^{\pm1},\dots,x_p^{\pm1},1),
\end{equation}
where $\oddrows\!\big((c^r)/\nu\big)=\ell$ means that the number of
rows of odd length in the skew shape $(c^r)/\nu$ equals exactly
$\ell$.
Use of this formula in \eqref{eq:sum-ortho} gives
\begin{equation} \label{eq:sum-schur2}
\det_{1\le s,t\le p}
\( q^{(s-\frac {1} {2})t}-q^{-(s-\frac {1} {2})t}\)
\sum _{\ell=0} ^{p}
s_{((H-p)^{p-\ell},(H-p-1)^\ell)}(q^{-p},\dots,q^{-1},1,q,\dots,q^p).
\end{equation}
Again, the hook-content formula \eqref{e7} applies and yields
\begin{multline*}
\det_{1\le s,t\le p}
\( q^{(s-\frac {1} {2})t}-q^{-(s-\frac {1} {2})t}\)
\sum _{\ell=0} ^{p}
q^{-(H-p)\binom {p+1}2+\binom {\ell+1}2}\frac {\prod _{s=1}
^{H-p-1}(q^{p+s+1};q)_p} {\prod _{s=1} ^{H-p}(q^{s};q)_p}\\
\times\frac
{(q^{H+\ell+1};q)_{p-\ell}\,(q^{\ell+1};q)_{p-\ell}\,(q^{H-p};q)_\ell}
{(q;q)_{p-\ell}}
\end{multline*}
for the expression \eqref{eq:sum-schur2}. Here we used the standard
notation for shifted $q$-factorials,
$(a;q)_k:=(1-a)(1-aq)\cdots(1-aq^{k-1})$, $k\ge1$, $(a;q)_0:=1$.
In terms of the standard basic hypergeometric notation
$${}_r\phi_s\!\left[\begin{matrix} a_1,\dots,a_r\\ b_1,\dots,b_s\end{matrix}; q,
z\right]=\sum _{\ell=0} ^{\infty}\frac {\poq{a_1}{\ell}\cdots\poq{a_r}{\ell}}
{\poq{q}{\ell}\poq{b_1}{\ell}\cdots\poq{b_s}{\ell}}\left((-1)^\ell q^{\binom
\ell2}\right)^{s-r+1}z^\ell\ ,$$
this can be written in the form
\begin{equation*}
\det_{1\le s,t\le p}
\( q^{(s-\frac {1} {2})t}-q^{-(s-\frac {1} {2})t}\)
q^{-(H-p)\binom {p+1}2}\bigg(\prod _{s=1} ^{H-p}\frac
{(q^{p+s+1};q)_p} {(q^{s};q)_p}\bigg)
{}_2\phi_1\!\[\begin{matrix} q^{H-p},q^{-p}\\q^{H+1}\end{matrix}; q,
-q^{p+1}\].
\end{equation*}
The determinant is again easily evaluated by means of \eqref{eq:sympl}.
On the other hand,
the $_2\phi_1$-series in the above expression can be summed with the help
of the $q$-analogue of Kummer's summation (see \cite[Appendix (II.9)]{GaRaAA}),
$$_2\phi_1\!\left[\begin{matrix} a,b\\aq/b\end{matrix};q,-q/b\right]=
\frac {(-q;q)_\infty\,(aq;q^2)_\infty\,(aq^2/b^2;q^2)_\infty}
{(-q/b;q)_\infty\,(aq/b;q)_\infty}\ .$$
Thus we have finally evaluated the sum in
\eqref{eq:vicious-with-2-asy-sum-even}. From there, it is then routine
to arrive at the expression \eqref{eq:starvicious-with-2even}.
\end{proof}

Clearly, if we specialize Theorem~\ref{thm:starvicious-with-2} to
$a_i=a+2i-2$, $i=1,2,\dots,p$, we obtain the asymptotics for stars
between two walls.

\begin{Corollary} \label{thm:star-with-2}
Let $a$ be an integer with $0\le a\le h-2p+2$.
The number of stars with $p$ branches of length $m$, the
$i$-th branch starting from $A_i=(0,a+2i-2)$,
$i=1,2,\dots,p$,
which do not go below the $x$-axis nor above the line $y=h$,
is asymptotically
\begin{multline} \label{eq:star-with-2even}
\frac {4^{p^2}} {(h+2)^p}  \bigg(2^p \prod _{s=1} ^{p}\cos \frac {s \pi}
{h+2}\bigg)^m
 \prod _{1\le s<t\le p} ^{}\sin^2 \frac {\pi  (t-s)} {h+2} \\
\times
 \prod _{1\le s\le t\le p} ^{}\sin \frac {\pi  (a+t+s-1)} {h+2} \cdot
  \sin \frac {\pi  (t+s)} {h+2}
 \prod _{s=0} ^{p}\prod _{t=1} ^{p}\frac {\sin \frac
   {\pi  (t-s+\fl{(h+2)/2})}
   {h+2}} {\sin \frac {\pi  (t-s+p)} {h+2}}\\
\times
 \prod _{s=1} ^{p}\frac {\sin \frac {\pi(s+\fl{(h+2)/2}-p)} {h+2}}
   {\sin \frac {\pi(2s+\fl{(h+2)/2}-p)} {h+2}},
\end{multline}
if $m+a$ is even, and
\begin{multline} \label{eq:star-with-2odd}
\frac {4^{p^2}} {(h+2)^p}  \bigg(2^p \prod _{s=1} ^{p}\cos \frac {s \pi}
{h+2}\bigg)^m
 \prod _{1\le s<t\le p} ^{}\sin^2 \frac {\pi  (t-s)} {h+2} \\
\times
 \prod _{1\le s\le t\le p} ^{}\sin \frac {\pi  (a+t+s-1)} {h+2}
\cdot   \sin \frac {\pi  (t+s)} {h+2}
 \prod _{s=0} ^{p}\prod _{t=1} ^{p}\frac {\sin \frac
   {\pi  (t-s+\fl{(h+1)/2})}
   {h+2}} {\sin \frac {\pi  (t-s+p)} {h+2}},
\end{multline}
if $m+a$ is odd.
\end{Corollary}

A different specialization
yields the asymptotics for $\infty$-friendly stars between two walls.

\begin{Corollary} \label{thm:infty-star-with-2}
Let $a$ be an integer with $0\le a\le h-2p+2$.
The number of $\infty$-friendly stars
in the TK model with $p$ branches of length $m$, the
$i$-th branch starting from $A_i=(0,a+2i-2)$,
$i=1,2,\dots,p$,
which do not go below the $x$-axis nor above the line $y=h$,
is asymptotically
\begin{multline} \label{eq:infty-star-with-2even}
\frac {4^{p^2}} {(h+2p)^p}  \bigg(2^p \prod _{s=1} ^{p}\cos \frac {s \pi}
{h+2p}\bigg)^m
 \prod _{1\le s<t\le p} ^{}\sin \frac {2\pi  (t-s)} {h+2p} \cdot
    \sin \frac {\pi  (t-s)} {h+2p}\\
\times
 \prod _{1\le s\le t\le p} ^{}\sin \frac {\pi  (a+2t+2s-3)} {h+2p} \cdot
  \sin \frac {\pi  (t+s)} {h+2p}
 \prod _{s=0} ^{p}\prod _{t=1} ^{p}\frac {\sin \frac
   {\pi  (t-s+\fl{(h+2p)/2})}
   {h+2p}} {\sin \frac {\pi  (t-s+p)} {h+2p}}\\
\times
 \prod _{s=1} ^{p}\frac {\sin \frac {\pi(s+\fl{(h+2p)/2}-p)} {h+2p}}
   {\sin \frac {\pi(2s+\fl{(h+2p)/2}-p)} {h+2p}},
\end{multline}
if $m+a$ is even, and
\begin{multline} \label{eq:infty-star-with-2odd}
\frac {4^{p^2}} {(h+2p)^p}  \bigg(2^p \prod _{s=1} ^{p}\cos \frac {s \pi}
{h+2p}\bigg)^m
 \prod _{1\le s<t\le p} ^{}\sin \frac {2\pi  (t-s)} {h+2p} \cdot
    \sin \frac {\pi  (t-s)} {h+2p}\\
\times
 \prod _{1\le s\le t\le p} ^{}\sin \frac {\pi  (a+2t+2s-3)} {h+2p}
\cdot   \sin \frac {\pi  (t+s)} {h+2p}
 \prod _{s=0} ^{p}\prod _{t=1} ^{p}\frac {\sin \frac
   {\pi  (t-s+\fl{(h+1)/2})}
   {h+2p}} {\sin \frac {\pi  (t-s+p)} {h+2p}},
\end{multline}
if $m+a$ is odd.
\end{Corollary}
\begin{proof}As in the proof of equation
\eqref{eq:infty-watermelons-with-2-asy}, we shift the $i$-th branch
up by $2(i-1)$ units. Thus we obtain a set of vicious walkers
with $p$ branches of length $m$, the
$i$-th branch starting from $A_i=(0,a+4i-4)$,
$i=1,2,\dots,p$,
which do not go below the $x$-axis nor above the line $y=h+2p-2$.
Hence, Theorem~\ref{thm:starvicious-with-2} with $a_i=a+4i-4$,
$i=1,2,\dots,p$, and $h$ replaced by $h+2p-2$ immediately gives the
desired asymptotics.
\end{proof}

As we noted in the case of watermelons, the restriction to stars
confined between two walls,
in contrast to stars without restriction and
with the restriction of {\em one} wall,
the asymptotics of (ordinary) stars
and $\infty$-friendly stars (compare the bases of the exponentials in
Corollaries~\ref{thm:star-with-2} and \ref{thm:infty-star-with-2})
is of a different order of
magnitude (except in the case of single branch stars, of course).
Hence, again, from these considerations, it is impossible to
conclude whether $n$-friendly stars restricted by two walls will have
the same order of magnitude as (ordinary) stars, or not.

As a final remark we mention that we could also have stated an
asymptotic formula the number of {\it all\/} possible vicious walkers
between two walls (i.e., with arbitrary starting {\it and\/} end points).
One would have to sum up the expression in Theorem~\ref{thm:starvicious-with-2}
over all $0\le a_1<a_2<\dots<a_p\le h$, which can be accomplished
in the same way as the summation over $0\le e_1<e_2<\dots<e_p\le h$
of the expression \eqref{eq:vicious-with-2-asy} in the proof of
Theorem~\ref{thm:starvicious-with-2}. We omit an explicit
statement for the sake of brevity.

\end{document}

